\documentclass[11pt,twoside]{article}  
\usepackage{adassconf}

\begin{document}   

%
%
%

\paperID{P1-1-17}

%
%
%
%

\title{VWhere: A Visual, Extensible `where' Command}

%
%
%

\author{Michael S. Noble}
\affil{Center for Space Research, Massachusetts Institute of Technology}

%
%

\contact{Michael S. Noble}
\email{mnoble@space.mit.edu}

%
%
%
%
%

\paindex{Noble, M.}

%
%

\authormark{Noble}

%
%



\begin{abstract}          
In this paper we describe {\itshape VWhere}, a S-Lang guilet which augments the
computational power of {\itshape where} with the point-click ease of a
Gtk-based visual interface.  The result is a new mechanism for interacting
with datasets, which unifies the constraint evaluation cycle implicit within
observational analysis and provides a number of compelling advantages
over traditional tool-based methods.
\end{abstract}

%
%
\section{The Problem: Exploratory Analysis}
The act of working raw data into a form from which its innate properties
may be more readily discerned, referred to here as {\itshape exploratory
analysis}, is of fundamental importance to scientific inquiry.  For example,
from an astrophysical dataset {\itshape \bfseries D}  one may wish to examine events within
some region $\mathcal{R}$, after time $\mathcal{T}$, but only if they originated
from detectors $\mathcal{I}$ and $\mathcal{J}$, and were above energy 
$\mathcal{E}_{0}$ but below energy $\mathcal{E}_{1}$.
Such constraint sets {\itshape \bfseries C} = \{$C_{0}, ..., C_{i}$\} are
derived by
iterating through ``{\itshape What If?}'' cycles, modeled in the large as
\begin{figure}[h]
\epsscale{.78}
\plotone{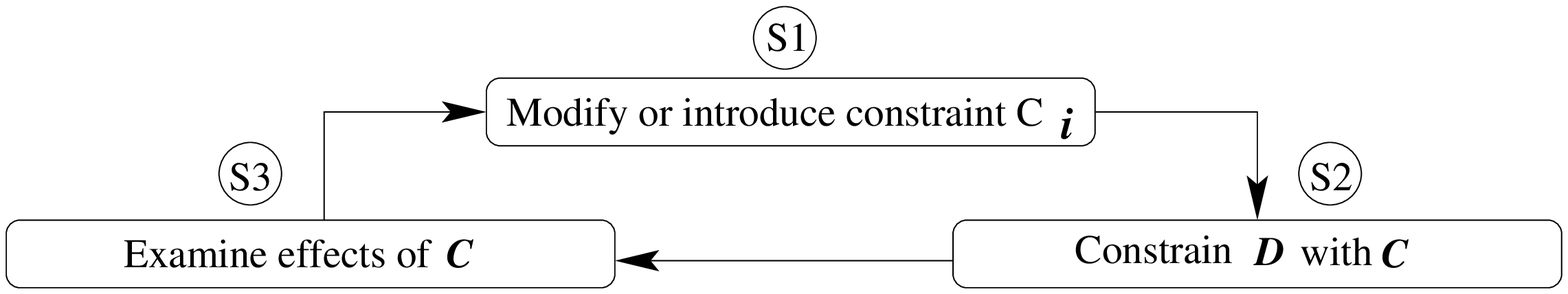}
\end{figure}\\
Stages 
{\hspace{0.05in}\begin{picture}(10,10) \put(3,3){\circle{13}}\put(-2.7,0){\small S1} \end{picture} \hspace{-0.10in}}\
and 
{\hspace{0.05in}\begin{picture}(10,10) \put(3,3){\circle{13}}\put(-2.7,0){\small S3} \end{picture} \hspace{-0.10in}}\
may involve numerical and visualization tasks, such as
fitting models and plotting residuals.  Stage
{\hspace{0.05in}\begin{picture}(10,10) \put(3,3){\circle{13}}\put(-2.7,0){\small S2} \end{picture} \hspace{-0.10in}}\
may be manifested
by applying filters to columns and images, or grouping, binning, et cetera.
This process plays out whether {\itshape \bfseries D} is a single observation or multiple
observations spanning a number of wavelengths and telescopes.
\subsection{A Traditional Solution: Command Line Tools}
Many existing astronomical analysis systems support data exploration via
some combination of command line tools and interactive applications.  The
tools in this model operate as file transformers ({\itshape file in ---$>$
file out}), and characterize the exploration process as one in which
{\hspace{0.05in}\begin{picture}(10,10) \put(3,3){\circle{13}}\put(-2.7,0){\small S1} \end{picture},
{\hspace{0.05in}\begin{picture}(10,10) \put(3,3){\circle{13}}\put(-2.7,0){\small S2} \end{picture}}, and
{\hspace{0.05in}\begin{picture}(10,10) \put(3,3){\circle{13}}\put(-2.7,0){\small S3} \end{picture}}
are performed by\\\\
   {\hspace*{0.3in} $\bullet$} multiple invocations of distinct programs (e.g.
	 \verb+fcopy/dmcopy/XSpec+)\\
   {\hspace*{0.3in} $\bullet$} requiring many passes over files in {\itshape \bfseries D} \\
   {\hspace*{0.3in} $\bullet$} during which a number of intermediate files are produced \\\\
\normalsize
While this approach certainly works, and provides bookkeeping advantages if
the tools record their operation (e.g. as comments in a FITS header), it is
clumsy for ad-hoc inquiry and generates file litter.  It is also needlessly
consumptive of
time and computational resources (at times enormously so, as corroborated
by Davis et al, \adassxiv).  Moreover, it is easy to mistakenly use the
filtering syntax of one system when executing tools from another.  Finally,
traditional tools are static in function, incapable of being selectively
extended by the end-user via the importation of a module or script.
\subsection{A New Solution: Interactive `where' Command}
To anyone who has analyzed data within array-oriented systems 
such as IDL(tm), CIAO/ISIS, or PyRAF, the {\itshape where} command
should need little introduction.
Couched in intuitive syntax, it provides a
powerful mechanism for filtering arrays via arbitrary expressions.  For
example, consider plotting from an event list all photons whose position
(specified by X, Y arrays) fell within a circle of radius 10 centered at
the origin
\begin{verbatim}
              isis> i = where( X^2 + Y^2 < 10);
              isis> plot(X[i], Y[i]);
\end{verbatim}
The {\itshape VWhere}\ guilet detailed herein augments this computational
power with a graphical interface that permits constraints to be specified,
manipulated, and evaluated visually.  In this model constraint sets are
represented as a combination of plots and the region filters cumulatively
applied to them.  Plots are generated from the axes of a dataset
{\itshape \bfseries D} resident in memory, or from new axes derived by
arbitrary analytic combinations of them, all without incurring any additional
I/O overhead or tool execution costs per iteration.
\section{Exploring Cygnus X-1}
Let us now walk through portions of a published research scenario, in which
{\itshape VWhere}\ was
used to correlate two observations (from Chandra and the RXTE All-Sky Monitor)
of black hole candidate Cygnus X-1. We begin by loading {\itshape VWhere}\ into ISIS, done explicitly here to underscore the fact that it is
\begin{verbatim}
              isis> require("vwhere");
\end{verbatim}
{\itshape not} linked directly into ISIS but is rather a standalone module
that can be loaded at runtime into {\itshape any} application which embeds
the S-Lang interpreter.
To maximize benefit to the community, while still promoting freedom of choice,
we've also begun to investigate means by which S-Lang modules may be
utilized within other interpreters, such as Tcl or Python.
After loading the data into a structure (not shown) we launch {\itshape VWhere}\
\begin{verbatim}
              isis> indices = vwhere(data);
\end{verbatim}
and generate a lightcurve by plotting 3 fields: {\tt time} versus
{\tt lchan + mchan}.
Our objective is to look for times where Cyg X-1 is emitting harder X-rays
during low count states.
\begin{figure}[h]
\epsscale{.9}
\plottwo{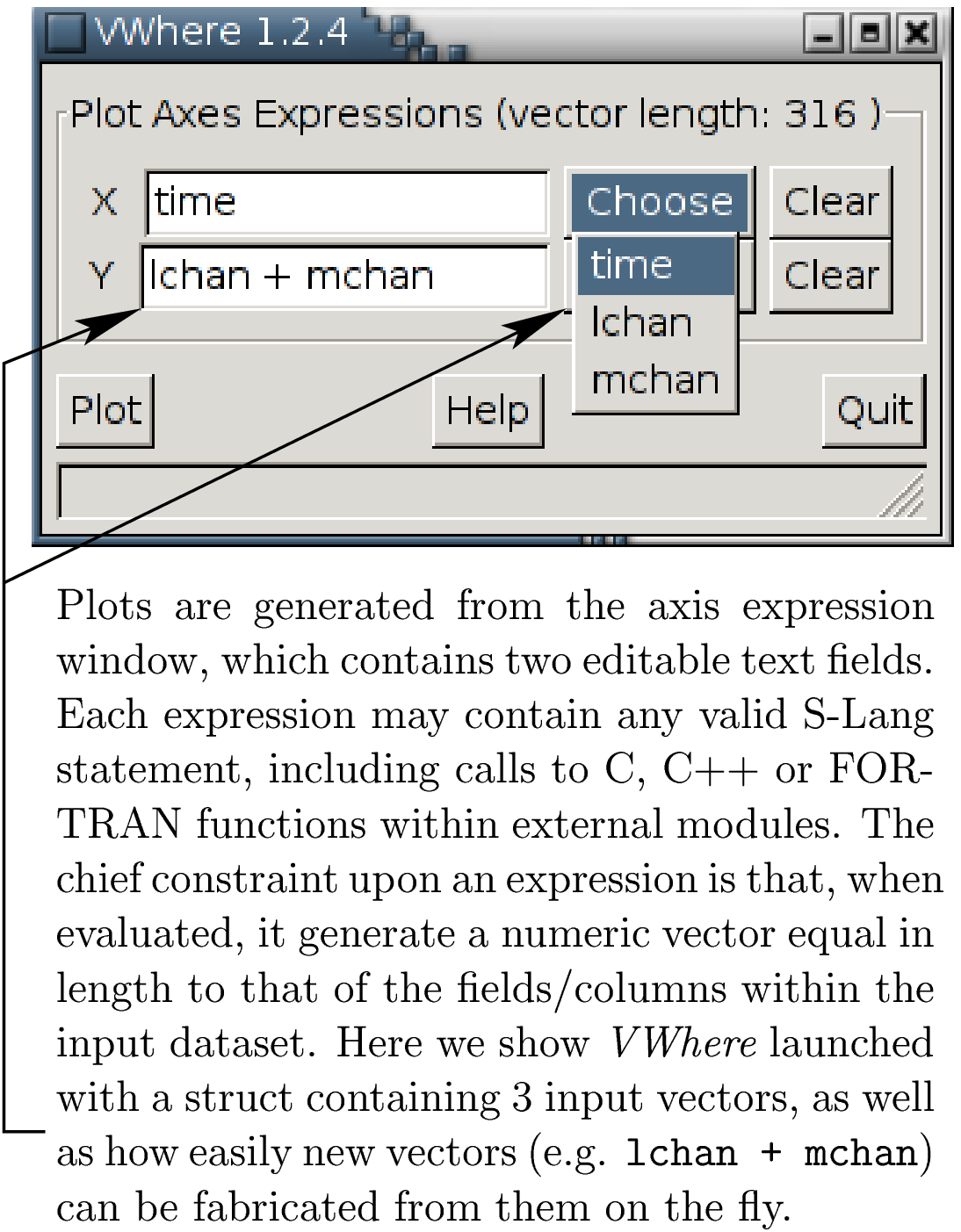} {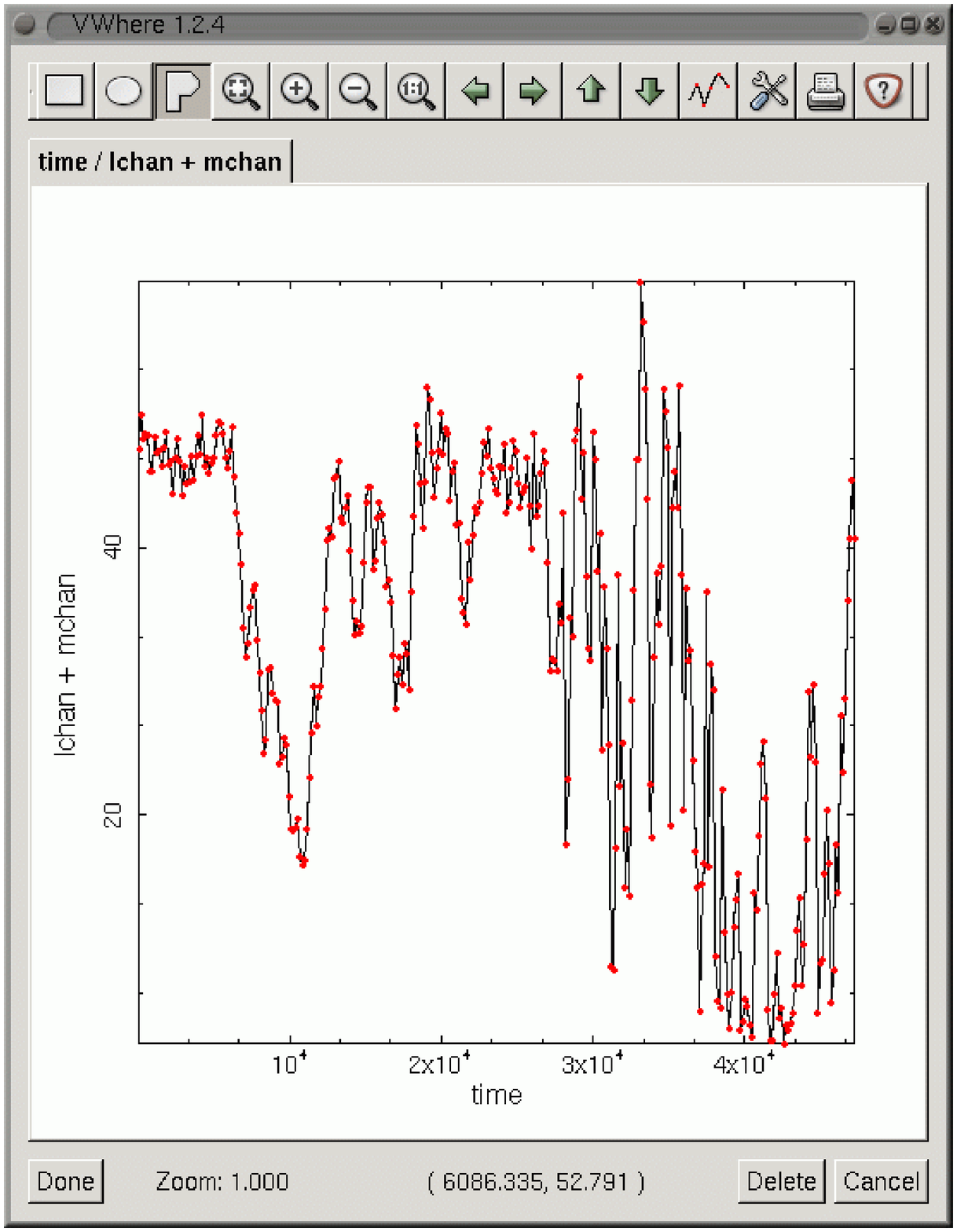}
\caption{A lightcurve for Cygnus X-1, and its axis expressions.}
\end{figure}
Each such plot is generated in its own tabbed subwindow, and may be
panned, zoomed, deleted, printed, or
cosmetically altered in many of the usual manners one would expect within a
plotting GUI.  Arbitrarily complex filters may then be constructed by laying
regions upon plots.  Pressing {\itshape Done} returns to the
caller a list of indices representing all points selected by the
applied filters (the same semantics as {\itshape where}).

Notice that the lightcurve above exhibits a good deal of variability.  The
color intensity diagram in the first panel of Figure \ref{fig2} shows that
as count rate drops (lower X value) CygX-1 emission is getting harder (smaller
Y value), except for a spike at the very lowest rates.  So we select points
within that spike, by applying a polygonal region, and look at the lightcurve
again in the second panel.  The points filtered away are drawn
grayed out, revealing that the hard, low count-rate selection comes from a
deep dip near the end of the lightcurve.
\begin{figure}
\epsscale{.80}
\plottwo{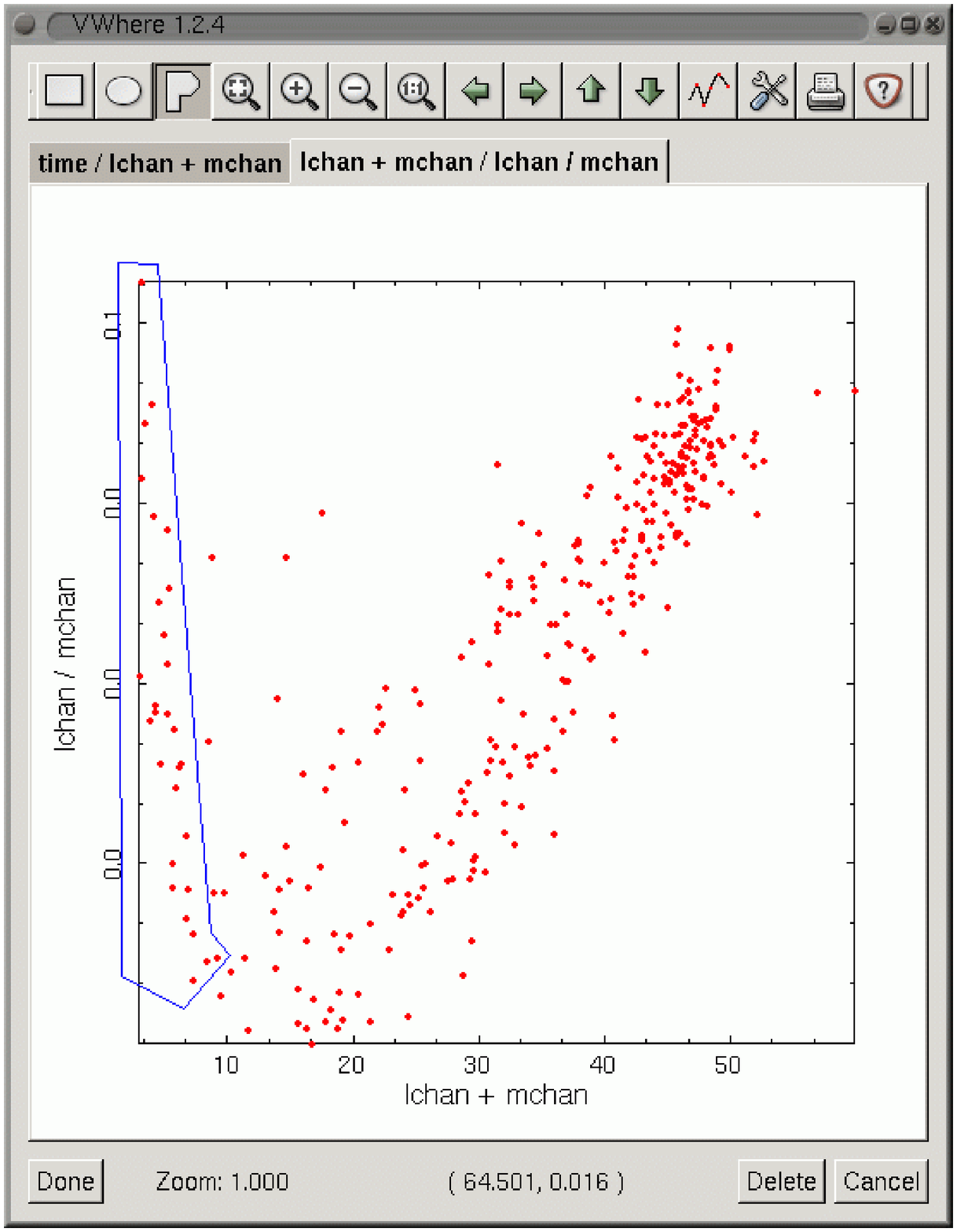} {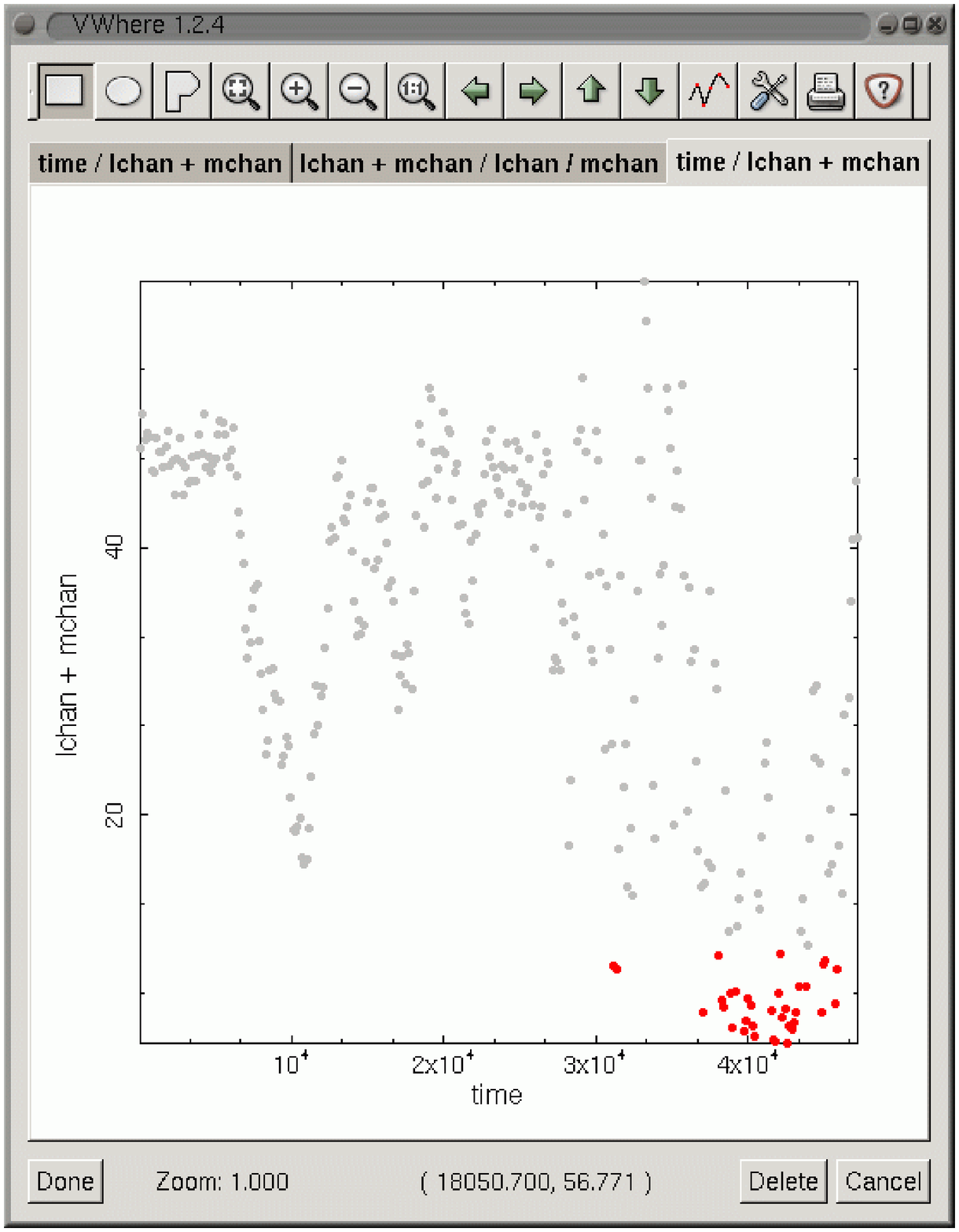}
\caption{A polygon filter applied to the color intensity diagram, and its incremental effect upon the original lightcurve.}
\label{fig2}
\end{figure}
\section{Benefit Analysis}
\subsection{Unifying the Constraint Cycle}
This example demonstrates how {\itshape incremental filtering} in
{\itshape VWhere}\ unifies stages
{\hspace*{0.05in}\begin{picture}(10,10) \put(3,3){\circle{13}}\put(-2.7,0){\small S1} \end{picture},
{\hspace{0.05in}\begin{picture}(10,10) \put(3,3){\circle{13}}\put(-2.7,0){\small S2} \end{picture}}, and
{\hspace{0.05in}\begin{picture}(10,10) \put(3,3){\circle{13}}\put(-2.7,0){\small S3} \end{picture}}
of the constraint evaluation cycle, dramatically shrinking the time and effort
required to discern subtle patterns within data, and without any package-specific filtering syntax.  With {\itshape VWhere}\ it is instantly clear how cuts
applied to portions of a dataset, or new axes derived from it, affect other
portions.
As a result, the process of constructing constraint sets is considerably more
fluid.  It is also far more powerful, since by virtue of the extensibility
endowed by S-Lang {\itshape VWhere}\ may import a wide range of C, C++, or
FORTRAN modules, and rather easily using the 
\htmladdnormallinkfoot{SLIRP code generator}{http://space.mit.edu/cxc/software/slang/modules/slirp/}. This is simply not feasible with traditional tools.
Finally, since {\itshape VWhere}\ operates directly upon arrays already
loaded within its host application, memory is conserved while I/O overhead
and runtime performance are improved by many factors.

\subsection{Real Modularity}{
Drawing from our experience as both user and developer of major
astronomical packages, we note a tendency in extant systems whereby
algorithms or libraries are internally coupled in ways which sharply 
curtail their use elsewhere.  Such systems, while open in the superficial
sense that sourcecode is
publicly available, are in a deeper sense closed by their complex web
of internal dependencies, which accrete to form a monolithic structure
impenetrable to the practicing scientist and difficult to navigate even
for professional programmers.

Potential users must confront the prospect of downloading, sifting
through, and picking apart tens or hundreds of thousands of lines of
software simply to obtain a small library or algorithm.  This effectively
prevents the use of novel {\itshape portions} of a system apart from its
{\itshape whole}.
Conversely, tightly coupled architectures betray a distinct evolutionary
disadvantage, in that they also tend to inhibit the incorporation of
software developed externally to the system.

{\itshape VWhere} is one of the earliest examples of our response to
this problem.  Our methodology shuns monolithic constructions in favor of
modular, extensible components that are orthogonal in function, and which
may either be woven together to form larger systems or used completely
standalone.  Other fruits of this effort are described on our
\htmladdnormallinkfoot{modules page}{http://space.mit.edu/cxc/software/slang/modules/} and in Davis et al (\adassxiv).

\subsection{SLgtk}
As argued elsewhere (e.g. Primini \& Noble, \adassxiv), S-Lang is an excellent
open-source language for scientific scripting, particularly for those with
FORTRAN, C, or IDL experience.  SLgtk augments its core strengths
(e.g. built-in support for array-based numerics) with an importable module
that makes it possible to construct sophisticated graphical interfaces
directly from the S-Lang interpreter.  Most of SLgtk is generated
automatically by SLIRP.

\subsection{Guilets}
A driving force behind the development of SLgtk has been the notion of 
the {\em guilet}, by which we mean visual software of a small, scriptable
nature which may be easily embedded within other applications, even those
with a primarily textual interface.  This fosters the use of graphical
interfaces where they are appropriate and beneficial, without committing
the entire application to such.

In code size and development cycle the typical guilet is considerably smaller
than traditional GUIs (applications designed with a primarily visual interface,
coded in a compiled language and explicitly utilizing low-level toolkits
such as Xt, Motif, or the Windows Foundation Classes).  Moreover, since
guilets are scripts they are more amenable to changing requirements, rapid
feature evolution, and even
customization by the end-user.  This can be vital to scientists\\
\begin{itemize}
\item interested in advanced analysis not yet supported by existing systems
\item or with proprietary periods who are unable to wait months for upgrades
\item or who find it advantageous to concoct working prototypes for
developers, to either supplement or supplant paper specifications\\
\end{itemize}
In point of fact, {\itshape VWhere} provides modern analysis and plotting
capability in only 1650 lines of scripts, and has been enhanced on numerous
occasions with turnaround times as short as 2 hours.
\section{Summary}
This paper describes {\em VWhere}, which extends the S-Lang {\em where}
command with a Gtk-based visual interface.  We have argued that by
integrating the specification, accumulation, and visualization of
analytical constraints {\em VWhere} presents a facile mechanism for
exploring data, one which eclipses tool-based methods for such,
particularly in terms of performance, extensibility, and ease of use.
\acknowledgments
This work was supported in part by the Chandra X-Ray Center contract
SV1-61010 from the Smithsonian Institution.  The author would like to thank
colleagues at MIT and the Harvard-Smithsonian CFA (especially Michael Nowak),
as well as those within the open source community, for various contributions
and commentary.

\end{document}